**OVERVIEW OF INITIAL RESULTS FROM THE RECONNAISSANCE FLYBY OF A KUIPER BELT PLANETESIMAL: 2014 MU$_{69}$** S.A. Stern[1], J.R. Spencer[1], H.A. Weaver[2], C.B. Olkin[1], J.M. Moore[3], W.M. Grundy[4], G.R. Gladstone[5], W.B. McKinnon[6], D.P. Cruikshank[3], L.A. Young[1], H.A. Elliott[5], A.J. Verbiscer[7], J.W. Parker[1], and the New Horizons Team. [1]Southwest Research Institute, 1050 Walnut St. Suite 300, Boulder, CO 80302 (astern@boulder.swri.edu), [2]Johns Hopkins University Applied Physics Laboratory, Laurel, MD 20723, [3]NASA Ames Research Center, Moffett Field, CA 94035. [4]Lowell Observatory, 1400 West Mars Hill Road, Flagstaff, AZ 86001, [5]Southwest Research Institute, 6220 Culebra Road, San Antonio, TX, 78238, [6]Dept. Earth and Planetary Sciences, Washington University, St. Louis, MO 63130. [7]Department of Astronomy, University of Virginia, Charlottesville, VA 22904.

**Introduction:** The centerpiece objective of NASA's New Horizons first Kuiper Extended Mission (KEM-1) was the close flyby of the Kuiper Belt Object (KBO) 2014 MU$_{69}$, nicknamed Ultima Thule [1]. On 1 Jan 2019 this flyby culminated, allowing New Horizons to make the first close up observations of a small KBO. Trajectory reconstruction indicates the spacecraft approach to within 3,500±34 km of MU$_{69}$ at 05:33:19±00:00:10 UT.

Here we summarize the earliest results obtained from that successful flyby. At the time of this abstract's submission, only 4 days of data downlink from the flyby were available. Well over an order of magnitude more data will be downlinked by the time of this LPSC meeting in 2019 March. Therefore many additional results not available at the time of this abstract submission will be presented in this review talk.

**Objectives:** 2014 MU$_{69}$ was discovered by members of the New Horizons team using the Hubble Space Telescope in 2014 [2]. MU$_{69}$'s orbit identifies it as a cold classical KBO. This means it has likely been resident near its current heliocentric distance (43 AU) and cold conditions continuously for the past ~4.5 Gyrs. This, combined with its small size, which prevents it from maintaining a strong internal geologic engine to the present, combine to make MU$_{69}$ the most primitive body ever studied by any planetary spacecraft.

Other than its orbital parameters and brightness, the only information known about Ultima Thule prior to observations by New Horizons were its red color [3], an approximate size (25-30 km diameter) and a shape profile at the time of a 17 July 2017 stellar occultation [2], and a visible albedo estimate (~0.1) from the occultation-measured size [2].

Key objectives of the flyby of MU$_{69}$ [1] were: to obtain panchromatic imaging, visible/near-IR wavelength four-color imaging, and stereo imaging of MU$_{69}$'s surface; to map Ultima Thule's surface composition across the 1.25-2.5 µm IR; and to make sensitive searches for both rings and satellites orbiting Ultima Thule and any gas or particulate coma. Secondary objectives included day and night side disk-integrated average brightness temperature measurements at a wavelength of 4 cm, an attempt to measure Ultima Thule's bistatic radar reflectivity at 4 cm, a search for dust impacts onto the spacecraft near Ultima Thule, and a search for detectable plasma interactions with the solar wind. Most of these objectives require data not yet on Earth before they can be written about. However, some data, comprising well under 1% of the total stored aboard New Horizons, is presently on the ground. We report on those data here.

**Initial Results.** New Horizons revealed MU$_{69}$ to be a bi-lobate contact binary that appears to have merged at low speed. In addition to being the most primitive solar system object ever to be explored in situ by any spacecraft, MU$_{69}$ thus also becomes the first primordial contact binary explored by spacecraft. The appearance and contact binary nature of this object is consistent with it being a relic planetesimal possibly created by pebble cloud collapse. How MU$_{69}$'s two lobes merged, how gently, and how much angular momentum was lost prior to final contact are puzzles to be solved as more data are returned and detailed modeling can be undertaken.

MU$_{69}$'s panchromatic and color appearances are shown in Figures 1 and 2, respectively.

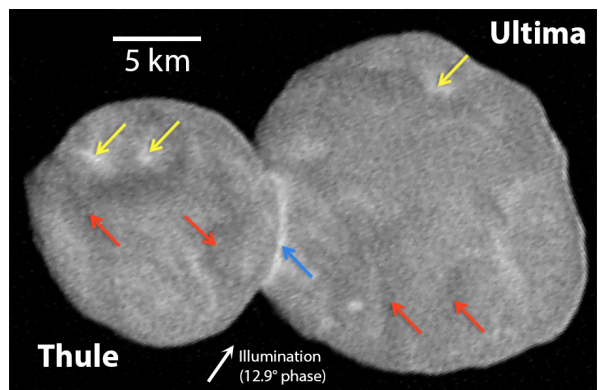

**Figure 1:** Ultima Thule as seen by the LORRI imager in the close approach CA04 observation (140 m/pixel), including relatively bright circular patches a few km wide (yellow arrows), darker regions up to several km wide (red arrows), a bright, cylindrically symmetric neck (blue arrow), and quasi-linear and arcuate bright features; note also the mottled appearance.

---

[*] Some feature names mentioned in this paper are now formalized while others are informal

The two lobes of $MU_{69}$ have very similar, quasi-spherical shape profiles in the available imaging. This near sphericity, if proven real with 3D shape modeling, could be an important clue to their formation.

For the purposes of discussion, $MU_{69}$'s larger and smaller lobes have been informally designated "Ultima" and "Thule," respectively, by the New Horizons team. Ultima and Thule have best-fit measured diameters of ~19.5 and ~14.2 km, respectively, with errors of only a few percent at this time. This, combined with its low amplitude lightcurve determined by New Horizons, suggests that the two lobes have ~2.6:1 volume, and for equal dnsities, mass ratios. Ultima Thule's overall major axis length is 34 km.

Ultima Thule's rotation period has been measured by New Horizons at 15.9±0.1 hours. How it slowed to this relatively long period after merging is another puzzle to be solved.

Although a mottled appearance and limb topography (amplitudes >1 km for Ultima and ~0.5 km for Thule) were resolved on each lobe, geological features are not clearly identifiable in the currently available images, which are restricted to a limited range of low (11-13°) solar phase angles. However, the significant limb topography variations detected suggest a mechanically strong "crustal zone" and/or bulk interior, perhaps composed of some combination of $H_2O$-ice, $CO_2$-ice, other ices, and refractory organics/rock.

Although the two lobes of Ultima Thule have closely similar reflectivities, significant albedo variations of 0.06-0.14 are seen on both lobes. The average I/F for the object as seen in Figure 1 is 0.09. Initial analysis reveals that this average albedo is similar to other cold classical KBOs. The most prominent albedo feature is a narrow (<300 m tall) bright, cylindrically symmetric neck where the lobes Ultima and Thule are joined. The origin of this remarkable feature is unclear; either endogenic (e.g., fine particle accumulation) and/or exogenic (e.g., lobe merger) possibilities may be the cause.

Ultima Thule's overall color displays a red slope of +24.3% per 1000 Å at 5500 Å. Additionally, the 1.25-2.5 μm near-infrared I/F varies between 0.17 and 0.19.

Little significant color variegation across either lobe has been detected as of this writing; however, the brighter neck region at the interface of the two lobes is slightly less red than the bulk color. Initial analysis reveals that Ultima Thule's color is similar to other cold classical KBOs (see also [3] with lower SNR HST color measurement than New Horizons). In comparing our Ultima Thule observations to Nix, we do not yet see strong evidence for the 2.0 μm $H_2O$-ice band; therefore the abundance of $H_2O$-ice on the surface is likely to be much less than observed by New Horizons on Pluto's satellites [4].

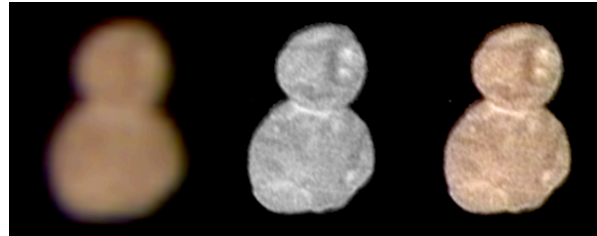

**Figure 2:** Left to right: Enhanced color image at 1.5 km/pixel resolution, panchromatic image at 140 m/pixel resolution, and enhanced color overlaid on the panchromatic image.

As expected, we found no evidence for a gas coma in initial inspections of UV airglow and solar appulse coma absorption datasets collected by the UV spectrograph on New Horizons, though the majority of those data are still to be downlinked. As also expected, no solar wind interaction or emitted pickup ions were detected in initial data analyses of the New Horizons plasma and charged particle spectrometer instruments, though the majority of those data are also still to be downlinked.

Regarding dust near Ultima Thule, New Horizons saw no evidence for backscattered coma light down to I/F~$5 \times 10^{-7}$ in imaging available by 4 Jan 2019. In addition, zero impacts onto the New Horizons dust counter were detected inside $MU_{69}$'s Hill sphere. And no satellites were detected down to 1.5 km diameter (under the assumption of an Ultima Thule-like reflectivity) outside of 1000 km from Ultima Thule. A search for closer rings or satellites has not yet been performed.

**References.** [1] Stern, S. A., Weaver, H.A, Spencer, J.R., and Elliot, H.A. (2018), *Space Science Reviews 214*, 76-99. [2] Buie, M.W. et al. (2018), DPS Meeting #50, id.509.06. [3] Benecchi, S., et al. (2019), *Icarus*, in press. [4] Cook, J., et al. (2018), *Icarus*, 315, 30-45.

**Acknowledgements**. This work was funded by NASA. We thank NASA, the Deep Space Network, KinetX Aerospace, JPL, the entire present and past New Horizons team, and the Gaia and HST missions for making the flyby of $MU_{69}$ successful.

[*] Some feature names mentioned in this paper are now formalized while others are informal